\documentclass[aps,twocolumn,showpacs]{revtex4}
\usepackage{amssymb}
\usepackage{amsfonts}
\usepackage{amsmath}
\usepackage{mathrsfs}
\usepackage{graphicx}

\begin{document}

\title{Inertial mass = gravitational mass, what about momentum?}

\author{Xiang-Song Chen}
\email{cxs@hust.edu.cn}

\affiliation{School of Physics and
MOE Key Laboratory of Fundamental Quantities Measurement,
Huazhong University of Science and Technology, Wuhan 430074, China}

\date{\today}

\begin{abstract}
It has been tested precisely that the inertial and
gravitational masses are equal. Here we reveal that the inertial
and gravitational momenta may differ. More generally, the inertial
and gravitational energy-momentum tensors may not coincide:
Einstein's general relativity requires the gravitational
energy-momentum tensor to be symmetric, but we show that a
symmetric inertial energy-momentum tensor would ruin the
concordance between conservations of quantized energy and charge.
The nonsymmetric feature of the inertial energy-momentum
tensor can be verified unambiguously by measuring the transverse
flux of a collimated spin-polarized electron beam, and leads
to a serious implication that the equivalence principle and Einstein's
gravitational theory cannot be both exact.
%04.20.Cv	Fundamental problems and general formalism
%04.80.Cc	Experimental tests of gravitational theories
\end{abstract}
\maketitle

Tremendous and continuous efforts have been devoted to testing
the equality of the inertial and gravitational masses, with no
difference found so far \cite{EP}. Since relativity requires that gravity
comes from not only the mass, but also the momentum and the
flow of energy and momentum, all of which round together to make
the energy-momentum tensor, the test of equivalence between
gravity and inertia should naturally be extended to the whole
energy-momentum tensor. In this paper, we show that due to its
many components, the energy-momentum tensor has great power in
potentially discriminating inertia from
gravity, both theoretically and experimentally.
For example, in high-energy physics there
is a wide-spread picture that gluons carry about half of the
nucleon momentum. This picture corresponds to the gravitational
momentum in Einstein's theory, while the inertial momentum
fraction of the gluon is probably only about 1/5 \cite{Chen09}.

The gravitational and inertial energy-momentum tensors
($T_{\rm g}^{\mu\nu}$ and $T_{\rm i}^{\mu\nu}$) are
relativistic extensions of the gravitational and inertial
masses ($m_{\rm g}$ and $m_{\rm i}$):
\begin{eqnarray}
\vec F =\frac {m_{\rm g}M_{\rm g}}{r^2} \hat {\vec r} &\Rightarrow&
R_{\mu\nu}-\frac 12 g_{\mu\nu} R=-8\pi T_{\rm g}^{\mu\nu},
\label{Tg} \\
\vec f =m_{\rm i} \vec a &\Rightarrow&
\left\{
\begin{array}{l}
f^\mu = \frac {d }{d\tau} P_{\rm i}^\mu  {\rm ~(for~a~particle)}\\
\\
f^\mu =\partial_\nu T_{\rm i}^{\nu\mu} {\rm ~(for~a~field)}
\end{array}
\right.
\end{eqnarray}
(In this paper we take the units with $c=\hbar=G=1$.)

As for $m_{\rm g}$ and $m_{\rm i}$, the relation between
$T_{\rm g}^{\mu\nu}$ and $T_{\rm i}^{\mu\nu}$ is not known
{\it a priori}, and has to be answered by experiment.
In Einstein's theory, $T_{\rm g}^{\mu\nu}$ is well restricted,
though it is very hard to measure the
gravitational effect of the components other than
$T_{\rm g}^{00}$ (the energy). In comparison,
$T_{\rm i}^{\mu\nu}$ has much better experimental
accessibility, and receives strong theoretical constrain
as well. The concrete knowledge of $T_{\rm i}^{\mu\nu}$
can have much to say about the validity of the
equivalence principle and Einstein's equation.

It is not unreasonable to suspect that $T_{\rm g}^{\mu\nu}$
and $T_{\rm i}^{\mu\nu}$ might differ. After all, in field theory
the conservation law alone does not give a unique expression
for the energy-momentum tensor. From N\"other's theorem,
one can derive the canonical expression from any covariant
Lagrangian density $\mathscr L(\psi^a, \partial_\mu \psi^a)$:
\begin{equation}
{\cal T} ^{\mu\nu}=
\frac {\partial \mathscr L}{\partial (\partial_\mu \psi^a)}
\partial^\nu \psi ^a -g^{\mu\nu} \mathscr L ,
~{\rm with}~\partial_\mu {\cal T}^{\mu\nu}=0.
\end{equation}
But one can always add to ${\cal T} ^{\mu\nu}$ a
total-divergence term
\begin{equation}
{\mathscr S}^{\mu\nu}= \partial_\rho B^{[\rho\mu]\nu}
\end{equation}
with $B^{[\rho\mu]\nu}$ antisymmetric in $[\rho\mu]$. It gives
$\partial_\mu {\mathscr S}^{\mu\nu}=0$, and, for a finite system,
$\int {\mathscr S}^{0\nu}dV=0$. Therefore, ${\mathscr S} ^{\mu\nu}$
changes neither the conservation law nor the conserved four-momentum.
By choosing a suitable ${\mathscr S} ^{\mu\nu}$, one can obtain a
symmetric tensor $\Theta ^{\mu\nu}=\Theta ^{\nu\mu}$ \cite{Wein95}.
Such tensor is what can fit into the Einstein's equation, the
left-hand side of which requires $T_{\rm g}^{\mu\nu}$ to be symmetric
and conserved in a flat space-time
(namely, when gravity is neglected):
\begin{equation}
T_{\rm g}^{\mu\nu}=T_{\rm g}^{\nu\mu},
~~\partial_\mu T_{\rm g}^{\mu\nu}=0.
\end{equation}

We are going to show the key point that any conserved and
symmetric energy-momentum tensor $\Theta ^{\mu\nu}$ leads to
a concrete and testable consequence, which is unacceptable
for the inertial energy-momentum tensor $T_{\rm i}^{\mu\nu}$
{\em when spin plays a role}.

As is well-known, the symmetric and conserved $\Theta ^{\mu\nu}$
can be used to construct a conserved angular momentum tensor:
\begin{equation}
{\mathscr M}^{\lambda\mu\nu}=x^\mu \Theta^{\lambda\nu}-x^\nu \Theta^{\lambda\mu},
~{\rm with}~\partial_\lambda {\mathscr M}^{\lambda\mu\nu}=0.
\end{equation}
A remarkable feature of such a construction is that the total
(spin+orbital) angular momentum density,
\begin{equation}
{\mathscr J}^k=\frac 12 \epsilon_{ijk} {\mathscr M}^{0ij},
\end{equation}
is given by an orbital-like expression:
\begin{equation}
\vec {\mathscr J}=\vec x \times \vec {\mathscr P}=\vec x \times \vec {\mathscr K},
\label{xK}
\end{equation}
where ${\mathscr P}^i= \Theta^{0i}$ and ${\mathscr K}^i=\Theta^{i0}$
are the momentum density and energy-flow density defined by the symmetric
energy-momentum tensor, which gives $\vec {\mathscr P} =\vec {\mathscr K}$.
(In general the two quantities can disagree.)

Simple as it looks, Eq. (\ref{xK}) (particularly the latter expression)
predicts a strong sum rule for experimental test.
Consider a collimated, polarized beam of spin-$s$ particles moving
in the $z$ direction (cf. Fig. 1 below). We have
\begin{equation}
\int (\vec x \times \vec {\mathscr K})^zdV=\int {\mathscr J}^zdV=
N s, \label{Jz}
\end{equation}
with $N$ the number of particles in the integrated volume.
This says that if the transverse flow of energy is measured locally,
one can integrate the result to tell whether or not the energy-momentum
tensor can be symmetric.

As we mentioned, it is certainly too hard to employ gravitational
measurement to test Eq. (\ref{Jz}), and thus to tell whether the
gravitational energy-momentum tensor $T_{\rm g}^{\mu\nu}$ can indeed
be symmetric. On the other hand, the inertial energy flow is in principle
measurable, therefore the symmetry of the inertial energy-momentum
tensor $T_{\rm i}^{\mu\nu}$ can be put into concrete test by
experiment.

One may think that such an experiment is not at all urgent or even
necessary, since $T_{\rm i}^{\mu\nu}$ must be symmetric should one believe
in both Einstein's equation and an exact equivalence between gravity and
inertia. However, we reveal an alerting fact that $T_{\rm i}^{\mu\nu}$
receives other theoretical constraint which contradicts Eq. (\ref{Jz}).

Consider the spin-polarized beam of free electrons moving in the $z$ direction,
with the same energy ${\cal E}$ for each electron.
Besides its energy and momentum, the electron has another conserved quantity,
the electric charge. Since the electrons are quantized in actual detection,
the flow of charge must be proportional to the flow of energy:
\begin{equation}
\frac {\vec K_{\rm i}}{\cal E}= \frac {\vec j}{e} =\vec n.
\label{n}
\end{equation}
Here $K_{\rm i}^i=T_{\rm i}^{i0}$ is the density of inertial energy flow,
$\vec j$ is the electric current and $\vec n$ is the flux density of
electron number.

Noticing that $\frac 12 \vec x\times \vec j=\vec \mu$ is the density of
electron magnetic moment, we have
\begin{eqnarray}
\int (\vec x \times \vec K_{\rm i})^z dV&= &
\frac {\cal E}{e}\int (\vec x \times \vec j)^z dV
=\frac {\cal E}{e} \int  2\mu^z dV\nonumber\\
&=& \frac {\cal E}{e}(N\cdot 2\frac e{\cal E} \cdot \frac 12) =N\cdot 1.
\label{N1}
\end{eqnarray}

However, should $T_{\rm i}^{\mu\nu}$ be symmetric, Eq. (\ref{Jz}) gives
\begin{equation}
\int (\vec x \times \vec K_{\rm i})^z dV=N\cdot \frac 12,\label{N2}
\end{equation}
which differs from Eq. (\ref{N1}) by a factor of two!

It is interesting and important to note that the canonical expression
can satisfy Eq. (\ref{n}) and hence reconcile conservations of
quantized energy and charge. For the electron, the canonical
energy-momentum tensor is easily derived to be,
\begin{equation}
{\cal T}^{\mu\nu}=\frac i2 \bar \psi \gamma^\mu \partial ^\nu \psi +h.c.,
\end{equation}
where $h.c.$ stands for hermitian conjugate.
This gives the canonical momentum and energy-flow densities
\begin{eqnarray}
{\cal P}^i &=&{\cal T}^{0i}=\frac i2\psi^\dagger \partial^i \psi +h.c., \\
{\cal K}^i &=&{\cal T}^{i0}=\frac i2 \bar \psi\gamma^i \partial^0\psi +h.c.
\rightarrow{\cal E}\bar \psi\gamma^i\psi.
\end{eqnarray}
Here the last expression applies to the electron state with fixed energy ${\cal E}$,
in which case we find ${\cal K}^i =\frac {\cal E}{e} j^i$ by comparing to the
electric current expression $j^\mu=e\bar\psi \gamma^\mu\psi$.

It is worthwhile to remind that in the canonical expression
$\vec x\times\vec {\cal P}=\vec {\cal L}$ is merely the orbital angular momentum.
For the spin-polarized electron beam with no orbital angular momentum, one finds
\begin{eqnarray}
\int (\vec x \times \vec {\cal P})^zdV&=&\int d^3x {\cal L}^z=0,\\
\int (\vec x \times \vec {\cal K})^zdV&=&\frac {\cal E}{e}\int (\vec x \times \vec j)^zdV=N\cdot 1.
\end{eqnarray}

In comparison, the symmetric energy-momentum tensor of the electron is
\begin{equation}
\Theta^{\mu\nu}=\frac i4 \bar \psi (\gamma^\mu \partial ^\nu +\gamma^\nu\partial^\mu)\psi +h.c.
\end{equation}
The corresponding momentum and energy-flow densities are
\begin{eqnarray}
{\mathscr P}^i &=&\Theta ^{0i}=\Theta ^{i0}={\mathscr K}^i =
\frac i4 \bar \psi (\gamma^0 \partial ^i +\gamma^i\partial^0)\psi +h.c. \nonumber \\
&=&\frac 12 {\cal P}^i +\frac 12  {\cal K}^i.
\end{eqnarray}
This shows clearly that if there is no orbital angular momentum one finds
\begin{equation}
\int \vec x \times \vec {\mathscr K}dV=
\int \vec x \times \vec {\mathscr P}dV=
\frac 12 \int \vec x \times \vec {\cal K}dV.
\end{equation}

Another interesting and important fact to note is that if the beam has
only orbital angular momentum $l$ and no spin polarization, we have
\begin{eqnarray}
\int (\vec x \times \vec {\cal P})^z dV&=&\int{\cal L}^z dV=N\cdot l,\\
\int (\vec x \times \vec {\cal K})^z dV&=&
\frac {\cal E}{e}\int (\vec x \times \vec j)^z dV=N\cdot l,
\label{Nl}\\
\int (\vec x \times \vec {\mathscr K})^z dV&=&
\int (\vec x \times \vec {\mathscr P})^z dV=N\cdot l.
\end{eqnarray}
Namely, orbital angular momentum is unable to discriminate the
symmetric energy-momentum tensor from the canonical one. A key difference
between Eqs. (\ref{N1}) and (\ref{Nl}) is that the gyromagnetic ratio is 1
for orbital magnetic moment but 2 for spin magnetic moment.

It is in order here to comment on the implication of our above observations.
First of all, as we showed vividly, the energy-momentum tensor does not
have the long-assumed arbitrariness; especially, its symmetry can be
tested unambiguously by experiment. On the theoretical side, Einstein's
equation dictates a symmetric gravitational energy-momentum tensor, while
in a quantum measurement the concordance between conservations of
quantized energy and electric charge refutes a symmetric inertial
energy-momentum tensor and favors the nonsymmetric canonical expression.
Apparently, one has to give up at least one long-cherished holy grail:
either the exact equivalence between gravity and inertia, or the beautiful
Einstein's theory, or the conservation law in quantum measurement.

We believe that it would not take too long before the experiment sheds some
light: The measurable difference we predict for the symmetric and
canonical energy-momentum tensors is surprisingly large. Should it be
confirmed that the inertial energy-momentum tensor is indeed nonsymmetric
(and thus in no conflict with the hardly questionable quantum conservation law),
then one has to infer that, either (i) the equivalence between gravity and
inertia is exact, thus $T_{\rm g}^{\mu\nu}$ must take the same nonsymmetric
expression of $T_{\rm i}^{\mu\nu}$, then the theory cannot be that of Einstein
for spin-polarized objects; or (ii) Einstein's equation is correct and thus
$T_{\rm g}^{\mu\nu}$ is symmetric, then the equivalence principle must be
violated when spin plays a role.

\begin{figure}
\includegraphics[width=0.45\textwidth]{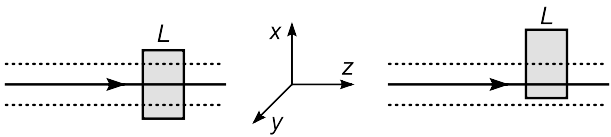}
\caption{\label{measure} A sketch of transverse flux measurement to
test the symmetry of the energy-momentum tensor. A cylindrically
symmetric beam goes along the $z$ axis. The thick solid line
is the symmetric axis of the beam, dashed lines mark the
beam width. The shaded box represents the detector window.}
\end{figure}

We discuss in some detail our proposed measurement as sketched in Fig. \ref{measure}.
For convenience, it is better to prepare a
cylindrically symmetric beam, then the integral in
Eqs. (\ref{N1}) or (\ref{N2}) simplifies to
\begin{equation}
\int (\vec x\times \vec K)^z dV=\int \rho K_\phi dV=2\pi L \int \rho^2 K_\phi d\rho,
\label{Kphi}
\end{equation}
where $(\rho,\phi,z)$ are the cylindrical coordinates, with the symmetric axis
of the beam chosen as the $z$ axis. The physical quantity to measure is $K_\phi$,
the density of transverse flux of energy. By preparing a beam with roughly the
same energy for each electron, the detector can just count the electron number.
The detector window is placed in the plane containing the $z$ axis. Since the
detector can only count the incoming electrons, the fluxes in both $\phi$ and
$-\phi$ directions need to be measured and subtracted to get the net flux
$K_\phi$. If the detector has good enough spatial resolution (say, 1/10 of the
beam radius), then the detector can be held fixed to record directly the
local flux density (see left panel of Fig. 1).
Otherwise, one has to gradually move the detector into
and out of the beam (see right panel of Fig. 1),
measuring the total flux at each step, and finally differentiating
with respect to the detector position to get the local flux.

To test the symmetry of the inertial energy-momentum tensor, the integrated result
in Eq. (\ref{Kphi}) is to be compared with $N$ or $N/2$ as predicted by
Eqs. (\ref{N1}) or (\ref{N2}). In actual practice, $N$ is the normalized
number of electrons in the integrated region, given by
\begin{equation}
N=I\cdot \frac Lv \cdot \xi \eta ,
\end{equation}
where $I$ is the beam intensity, $L$ is the detecting length along the beam direction,
$v$ is the electron velocity, $\xi$ is the counting efficiency, and $\eta$ is the
polarization rate.

Before closing this paper, we comment on the energy-momentum tensor
for other particles, particularly photons, composite and higher-spin
particles. For the photon, the symmetric and canonical energy-momentum
tensors are easily derived to be
\begin{eqnarray}
\Theta ^{\mu\nu} &=&F^{\mu\rho} F_\rho^{~\nu} +\frac 14 g^{\mu\nu} F^2 ,\\
{\cal T} ^{\mu\nu} &=&-F^{\mu\rho} \partial^\nu A _\rho +\frac 14 g^{\mu\nu} F^2 .
\end{eqnarray}
The canonical expression is gauge-dependent and hence often abandoned.
But recent intensive studies show that ${\cal T} ^{\mu\nu}$ can be revised
gauge-invariantly. (The issue is, though, highly tricky and controversial. We
refer interested readers to Refs. \cite{Chen09,Chen12} and references therein.)

The familiar Poynting vector $\vec E\times\vec B$ is the momentum and energy-flow
density of $\Theta ^{\mu\nu}$. Interestingly, for a free photon,
$\vec E\times\vec B$ is also the energy-flow density of ${\cal T} ^{\mu\nu}$
in the radiation gauge. Therefore, the type of measurement discussed above is
unable to distinguish the symmetric and canonical energy-momentum tensors for
the photon. To explore the symmetry of the photon energy-momentum tensor, one
has to measure the momentum-flow tensor, which is different for
$\Theta ^{\mu\nu}$ and ${\cal T} ^{\mu\nu}$. (The momentum density itself is
also different for $\Theta ^{\mu\nu}$ and ${\cal T} ^{\mu\nu}$,
but it is hard to conceive how to measure the momentum density).
Knowledge of the photon energy-momentum tensor would shed invaluable light on 
the gauge-dependence problem in canonical expressions, and offer a
deeper or even new understanding of the gauge symmetry \cite{Chen09,Chen12}.
(After the first version of this paper appeared, a major step forward was
made in Ref. \cite{Chen12b}, which discovered a clever way of analyzing 
the momentum flow, and shew that the symmetry of the photon 
energy-momentum tensor can actually be inferred from the {\em known} 
diffraction patterns of light carrying spin and orbital angular momentum,
respectively.)

Studying the energy-momentum tensors of composite and higher-spin particles
is also very interesting in its own right. If a beam of such particles
is used, then, unlike for the fundamental electrons or photons, we have
basically no prediction for the possible
outcome of measuring the transverse flux as in Fig. 1. It does not help
to express the energy-momentum tensor of a composite particle in terms
of its fundamental constituents, since it is the composite particle as
a whole that is quantized in a measurement. Actually performing the
type of measurement as in Fig. 1 for composite and high-spin particles
would help to study how their effective degrees of freedom can be
described by a quantum wavefunction.

This work is supported by the China NSF via Grants No.
11275077 and No. 11035003, and by the NCET Program of the China
Ministry of Education.

\end{document}